\documentclass[%
reprint,
 amsmath,amssymb,
 aps,
 prl,
]{revtex4-2}

\usepackage{graphicx}
\usepackage{dcolumn}
\usepackage{bm}
\usepackage[dvipsnames]{xcolor}
\usepackage{hyperref}
\usepackage{xr-hyper}
\usepackage{cleveref}
\makeatletter
\IfFileExists{supplementary_information.aux}{%
  \typeout{^^J*** Found the supplement .aux! ***^^J}%
}{%
  \typeout{^^J*** Did NOT find supplementary_information.aux ***^^J}%
} 
\makeatother
\begin{document}

\title{General perturbative framework for kinetics \\of rare transitions in 1-dimensional active particle systems }

\preprint{APS/123-QED}

\author{ Vito Seinen$^{1,6}$}
\author{ Peter G. Bolhuis$^2$ }%
\author{ Daan Crommelin$^{3,4}$ }%
\author{ Sara Jabbari Farouji$^{1}$ }%
\author{ Michel Mandjes$^{3,5}$ }%
\affiliation{%
$^1$Institute for Theoretical Physics, University of Amsterdam, Amsterdam, The Netherlands
}%
\affiliation{
$^2$Van ’t Hoff Institute for Molecular Sciences, University of Amsterdam, Amsterdam, The Netherlands
}%
\affiliation{%
$^3$Korteweg-de Vries Institute for Mathematics, University of Amsterdam, Amsterdam, The Netherlands
}%
\affiliation{%
$^4$Centrum Wiskunde \& Informatica, Amsterdam, The Netherlands
}%
\affiliation{%
$^5$Mathematical Institute, Leiden University, Leiden, The Netherlands
}%
\affiliation{%
$^6$Dutch Institute for Emergent Phenomena, Amsterdam, The Netherlands
}%

\date{ April 16, 2026 }

\begin{abstract} 
We present a theoretical framework that enables investigating rare transitions in a general model of an  active particle in an external potential, with the thermal Active Ornstein–Uhlenbeck Particle (AOUP) appearing as a special case. 
Using a projection-operator formalism, we compute transition rates perturbatively in two distinct asymptotic regimes. In the regime of small persistence times—where the activity evolves much faster than the particle’s position—integrating out the activity reproduces the rates previously reported in the literature. In the opposite regime of large persistence times, we instead integrate out the position and obtain the corresponding rates analytically.
Together, these asymptotic expansions uniquely specify a rational approximation that remains accurate across intermediate persistence times.
As a result, we obtain an analytic expression for the rate valid across all persistence times and activity strengths in the rare-event limit, which are in excellent agreement with numerical simulations. The presented framework applies to rare transitions in a broad class of driven systems. 
\end{abstract}

\pacs{123}


\maketitle

\noindent {\it Introduction.} 
Active matter, comprising systems that continuously inject energy to drive their own motion, has emerged as a flourishing research field in non-equilibrium statistical physics \cite{HowFarFromEquilibrium,Ramaswamy2010,Marchetti2013}. Particularly, colloids with a stochastic driving force have been popular as a minimal model for self-driven biological systems \cite{Bechinger2016,Needleman2017}, such as microorganisms, cells and molecular motors. Active systems exhibit many nonequilibrium effects that have no counterparts in passive systems, such as non-Boltzmann steady states and continuous entropy production \cite{HowFarFromEquilibrium,Statmech_AOUP}. Moreover, they undergo non-equilibrium phase transitions, such as Motility Induced Phase Separation (MIPS). In active model systems, as well as in the biological systems they aim to model, active barrier crossings contribute to nucleation in MIPS \cite{Cates2015}, analogously to passive systems \cite{ReviewKramer}. Transition rates for barrier crossings have been extensively studied for passive and close to equilibrium systems. In  such cases, one can often reduce a multidimensional problem to an effective 1D multi-well potential, allowing the use of Kramers’ theory as an effective approach to computing the transition rates for passive nucleation processes \cite{ReviewKramer}.

Progress in analytical calculation of transition (escape) rates for active systems, however, has been limited  to special cases. The small persistence time regime, where the active force changes its orientation and magnitude rapidly, has received a lot of attention. Here, perturbative calculations result in an effective temperature and effective potential that characterize the steady state \cite{LDT_transition,Martin_2021,Statmech_AOUP}. Calculating transition rates by applying Kramers' theory and taking into account the effective temperature contribution has shown good agreement with simulation results in the small persistence time regime \cite{Militaru_2021,correlated_escape}, where higher order corrections are not relevant. Beyond the small persistence time regime, although being further from equilibrium and more relevant for MIPS \cite{reviewMIPS}, results have been much more limited. 

For the Run-and-Tumble model, mean first passage times and therefore transition rates have been obtained exactly for arbitrary persistence times and parameter values \cite{Run_and_tumble}, while in the weak-noise, high-barrier limit the transition rates have been derived analytically using path integral methods \cite{TransitionActivePArticlePathintegrals}. 

For the Active Ornstein-Uhlenbeck Particle (AOUP) without thermal noise, the exponential scaling (i.e. neglecting subexponential prefactor) of the transition rates has been approximately calculated using large deviation theory, via expansions in both the low and large persistence-time limits \cite{LDT_transition}. However, as we will see, the inclusion of thermal noise strongly alters the behaviour, with transitions arising from the interplay between activity and thermal fluctuations.

In this work, we derive for the first time analytical expressions for the transition rates for a broad class of active particle models, regardless of persistence time and activity strength, in the rare event limit. We achieve this by reducing the eigenvalue equations, where the transition rate is given by the lowest eigenvalue in presence of an absorbing boundary, to one-dimensional effective equations. This is realized by integrating out the fast variable using a projection-operator formalism. We employ the Feshbach-Schur map, widely used in nuclear scattering and cold atom systems \cite{Feshbach,BookFeshbachCMT}. This yields an effective equation for the position variable in the regime of fast activity (small persistence time), and for the activity variable in the regime of slow activity (large persistence time). We derive asymptotic expansions of the transition rates in both regimes. For the AOUP case, we further construct a unique rational interpolation for the intermediate regime, using a two-point Pad\'e approximant, thereby obtaining the rate for all persistence times.

\begin{figure*}
    \centering
    \includegraphics[width=1.0\linewidth]{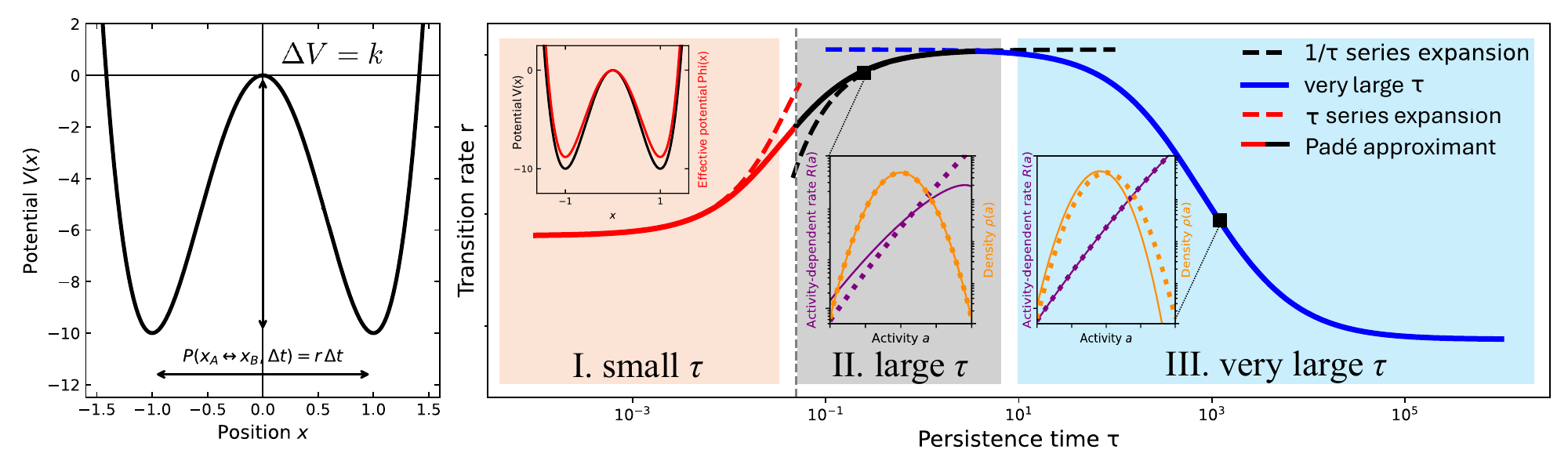}
    \caption{(Left) Double well potential $V(x) = 4k(-\frac{ x^{2} }{2} + \tfrac{ x^4 }{4} )$  depicted for $k=10$. 
    (Right) The calculated rate $r$ as a function of persistence time $\tau$. In the small $\tau$ regime (I), $\tau \ll 1/k$, the projection formalism gives an effective equation in the position variable $x$. In this regime, where the calculated rate is shown as a dashed red line, the rate is controlled by the quantity $\Phi(x)$ from Eq.\ \eqref{eq:Phi_x}, which one can interpret as a quasipotential in $x$ only. In the large $\tau$ regime (II and III), $\tau \gg 1/k$, 
    the projection formalism gives an effective equation in the activity variable $a$. In this regime, the rate is completely controlled by the well-dependent density $p_{\rm{a}}(a)$ and the activity-dependent transition rate $R(a)$, through the relation $r = \int^{\infty}_{-\infty} p_{\rm{a}}(a)R(a) da$, see Eq.\ \eqref{eq:totalrate_largetau}. The large $\tau$ limit has two subregimes. In the intermediate-to-large $\tau$ regime (II), the density $p_{\rm{a}}(a)$ is Gaussian, while $R(a)$ is $\tau$-dependent. In the very large $\tau$ subregime (III), $\tau$ is large enough that $R(a) \approx R_0(a)$, and is therefore $\tau$-independent, while $p_{\rm{a}}(a)$ becomes $\tau$-dependent. In the intermediate $\tau \sim 1/k$, a unique Padé approximant exists that is used to bridge the two asymptotic expansions I and II.} 
    \label{fig:Conceptual_figure}
\end{figure*}

To test our theory, we numerically simulate the Active Ornstein-Uhlenbeck particle in the example case of a double-well potential, see Fig.~\ref{fig:Conceptual_figure}(a), for intermediate and large barrier heights. We perform a parameter sweep in persistence time and activity strength, and compare to the theoretical predictions, where we consider barrier heights $\Delta V/k_BT = 7$ and $13$.

In the small-persistence-time regime, activity primarily increases the effective temperature, thereby enhancing the overall transition rate, in line with prior findings ~\cite{LDT_transition,Martin_2021,Statmech_AOUP}. Higher-order corrections include spatially dependent diffusion and effective forcing terms, which slightly reduce the transition rate.
In the large-persistence-time regime, particles whose activity happens to point toward the opposite well have a substantially increased probability of crossing the barrier. This probability is maximized when the persistence time is large compared to the local well relaxation time. However, as the persistence time increases further and becomes comparable to the mean first passage time $1/r$, the overall transition rate decreases again. This occurs because particles that have crossed must typically wait a time on the order of the persistence time before the activity reorients toward the opposite well, leading to non-exponential first-passage-time distributions.

We note that although our theory has been developed for active matter systems, it is also applicable to other one way driven stochastic systems where rare transitions occur \cite{Climate_tipping_paper} \cite{Bistable_neurons_paper}, as long as the conservative force of the driven variable is multi-stable. 
 
\noindent {\it Generalized model.}
We consider the system of coupled overdamped Langevin equations  describing the dynamics of the (dimensionless) position $x$ (in the overdamped limit) and activity $a$
\begin{subequations}
\begin{align}
    \dot{x} &= k F(x)  - \lambda F_{a}(x,a)+ \sqrt{2}\eta_{\rm{x}}(t),  \\
    \dot{a} &= - \frac{1}{ \tau } v'(a) + \sqrt{ \frac{ 2 }{ \tau } } \eta_{\rm{a}}(t).
\end{align}
\label{eq:generalSDE}
\end{subequations}
Here, $k F(x)= - V'(x)$ denotes the conservative force with multiple wells, $\eta_{\rm{x}}$ and $\eta_{\rm{a}}$ are independent Gaussian white noise processes with unit variance, and $v'(a)$ gives the dynamics of the activity variable $a$.
$\lambda$ gives the strength of the active force, 
and the persistence time $\tau$ is the timescale at which the activity changes. $F_a(x,a)$ is a generalized active force, which can be spatially dependent, and can be taken to be $F_{a} = a$ for the AOUP, with $v'(a) = a$. To retrieve the ABP, one chooses $F_{a} = \cos(a)$ and $v'(a) = 0$.
The corresponding dimensional variables (denoted by $^*$) are related to the dimensionless ones via
$x^{*} := L x$, $\tau^* := \frac{ \gamma L^{2} }{  k_B T } \tau$,  $k^*:= \frac{ k_B T }{ L } k$, $\lambda^* := \frac{k_B T }{ L } \lambda $ and $t^{*} := \frac{\gamma L^{2}}{k_{B} T} \, t$, see the Supplementary Material (SM Sec.\ I).
Here, the characteristic length scale $L$ gives the distance between the well and origin, the thermal energy $k_B T$ sets the energy scale, 
and $\gamma$ is the friction coefficient defining the diffusive timescale $\gamma L^2/(k_B T)$.
We specifically apply our theory to a double well potential $V(x) = 4 k (-\frac{ x^2 }{2} + \frac{x^4}{4})$, the barrier height then equals $k$.

\noindent {\it Fokker-Planck equation.} 
To obtain the transition (escape) rate, we turn to the Fokker-Planck (FP) equation corresponding to Eq. \eqref{eq:generalSDE}. The transition rate over a barrier can be found by imposing an absorbing boundary condition at a point $x_{b}$ beyond the barrier,{\it i.e.}, $p(x_{b},a) = 0$, where $x_b$ is chosen far enough over the barrier such that recrossings are unlikely, but before the minimum of the opposing well. In the high barrier limit $\Delta V = k \gg 1$, a particle near a well relaxes into the quasi-steady state, such that the solution to the time-dependent FP equation becomes $p(x,a,t) = e^{-rt} p_{ \rm{qss} }(x,a)$. Then, the time-dependent FP equation results in $\mathcal{L} p = \partial_{t}p = -r p$. Therefore, finding the transition rate $r$ is equivalent to solving the lowest eigenvalue $r$ in the equation
\begin{align}
    \mathcal{L} | p \rangle &= \left( \mathcal{L}_{x} + \tfrac{1}{\tau} \mathcal{L}_{a} \right) | p \rangle = -r | p \rangle, \label{eq:eigenvalueequation}\\
    \mathcal{L}_{x} \cdot &= - \partial_{x} ([ k F(x) + \lambda F_{a}(x,a)] \cdot ) + \partial_{xx} \cdot \\
    \mathcal{L}_{a} \cdot &= \partial_{a} (v'(a) \cdot ) + \partial_{aa} \cdot
\end{align} 
Here, $\mathcal{L}_{x}$ is the FP-operator acting on $x$, while $\mathcal{L}_{a}$ acts on $a$. We write the probability density $p = p(x,a)$ and henceforth all functions in bra-ket notation. 
As Eq.\ \eqref{eq:eigenvalueequation} can not be solved analytically, we resort to approximation.

\noindent {\it Projection formalism for timescale separation between position and activity. }  
Our approach to obtaining analytical expressions for the transition rate $r$ is to reduce the problem to an effective one-dimensional description. 
We carry this out in the limits of small and large $\tau$, where in each case we integrate out the fast variable using projection operators. Locally the wells are harmonic, such that the timescale at which the variable $x$ equilibrates scales as $\frac{1}{k}$ \cite{gardiner}, while the timescale at which the variable $a$ equilibrates is set by $\tau$. The small $\tau$ regime, where $a$ equilibrates much faster than $x$, is characterized by the condition $\tau \ll \frac{1}{k}$, so that $k \tau$ is a small parameter in this regime. The opposite large $\tau$ regime is characterized by the condition $\frac{1}{k} \ll \tau$, where $\frac{1}{k \tau}$ is a small parameter.
We reduce the effective dimensionality of our problem by introducing projection operators, projecting the dynamics associated with the fast variable, either $x$ or $a$, onto the slower variable. This projection operator is defined as $P := \lim_{t \rightarrow \infty} e^{ ( \mathcal{L}_{f} - l_{0}) t } $, where $\mathcal{L}_{f}$ is the operator corresponding to the fast variable and $l_{0}$ is its principal eigenvalue. In the general case $P$ takes the form $P = | \rho_{0} \rangle \langle \phi_{0}|$, where $| \rho_{0} \rangle$ is the principal eigenfunction solution to $\mathcal{L}_{f} | \rho_{0} \rangle = l_{0} | \rho_{0} \rangle $ and $| \phi_{0} \rangle$ is the principal eigenfunction solution to $\mathcal{L}_{f}^{\dagger} | \phi_{0} \rangle = l_{0} | \phi_{0} \rangle $, where $\mathcal{L}_{f}^{\dagger}$ denotes the backwards FP operator.
In the absence of an absorbing boundary for the 
fast variable $l_0=0$, the projection operator reduces to the widely used form $ \lim_{t \rightarrow \infty} e^{\mathcal{L}_{f}t}$ \cite{gardiner}. Defining the orthogonal  projection operator $Q = I - P$, one retains all information of any state $| \psi \rangle$ through the relation $|\psi \rangle= I |\psi \rangle = P | \psi \rangle + Q | \psi \rangle$. The following step is to express the state of interest $| p \rangle$ from Eq.\ \eqref{eq:eigenvalueequation} in terms of $P |p \rangle$ alone, which can be done exactly by making use of the Feshbach-Schur map \cite{Feshbach}, which allows us to express $Q | p \rangle$ as an operator acting on the state $P |p\rangle$ alone. We obtain (see  Sup. Sec.\ II ), 
\begin{subequations} 
\begin{align}
     & \langle \phi_{0} | \mathcal{L} P | p \rangle - \langle \phi_0 | \mathcal{M} P | p \rangle = - r \langle \phi_{0} | P | p \rangle, \label{eq:generalreducedequation} \\
     & \mathcal{M} := P \mathcal{L} Q \big( Q ( \mathcal{L} + r) Q \big)^{-1} Q \mathcal{L} P.
\end{align}
\end{subequations}
Eq.\ \eqref{eq:generalreducedequation} is a general result for projection operators and is used in nuclear \cite{Feshbach} and condensed matter physics \cite{BookFeshbachCMT}. Taking the inner product with the left principal eigenfunction $| \phi_0 \rangle$ results in an integration over the fast variable, so that Eq.~\eqref{eq:generalreducedequation} reduces to a one-dimensional equation in the slow variable. 
Importantly, $\mathcal{M}$ can be expanded in powers of $k \tau$ in the small $\tau$ regime, and $\frac{1}{k \tau}$ in the large $\tau$ regime. For convenience, we will expand in powers of $\tau$ or $1/\tau$, where the factors involving $k$ will be absorbed into the coefficients. 

\noindent {\it The small persistence time regime: $ \tau \ll 1/k $.} In the small $\tau$ regime, the fast variable is the activity $a$, such that $\mathcal{L}_{f} = \frac{1}{\tau} \mathcal{L}_{a}$. Since there is no absorbing boundary condition in the active variable $a$, we have that $l_{0} = 0$, $ | \rho_{0} \rangle = \frac{1}{Z_{a}} e^{-v(a)}$ and $|\phi_{0}\rangle = 1$. In this limit, we find that Eq. \eqref{eq:generalreducedequation} yields an effective FP equation for the slow variable $x$, see SM Sec.\ III, given by 
\begin{equation}
    - \partial_{x} \big(F_{ \rm{eff} }(x) \, p_{\rm{x}}(x) \big) + \partial_{xx} \big( D_{ \rm{eff} }(x) \, p_{\rm{x}}(x) \big) = -r p_{\rm{x}}(x).
    \label{eq:eigenvalueproblem_xspace}
\end{equation}
Here, $p_{\rm{x}}(x) := \langle \phi_0 | P | p \rangle = \int^{\infty}_{-\infty} p(x,a) da$. Expanding $\langle \phi_0 | \mathcal{M} P | p \rangle $ in powers of $\tau$ results in additions to the effective drift and diffusion. We now apply this formalism to the AOUP case, where $v(a) = a^2/2$ and obtain to second order in $\tau$, the effective diffusion $D_{\text{eff}}$ equals 
\begin{align}
    D_{ \rm{eff} }(x) & = 1 + \tau \lambda^{2} + \tau^{2} \lambda^{2} F'(x) + O \left( \tau^{3} \right),
    \label{eq:Deff}
\end{align} 
with no contributions to the effective force $F_{\rm{eff}}$, i.e. $F_{\rm{eff}} = F + O(\tau^{3})$. One can extend calculations to third order yielding $O(\tau^{3})$ contributions to both the force and diffusion. We neglect these because of the high precision of the second order rational approximation (discussed further down). 
From the effective drifts and diffusion coefficients, it is straightforward to find the transition rate. It is a standard result \cite{ReviewKramer} that for any given $D_{\rm{eff}}(x)$ and $F_{\rm{eff}}(x)$ the transition rate can be expressed as 
\begin{subequations}
\begin{align}
    r &= \frac{D_{ \text{eff} }(x_{\text{max}})}{2 \pi}\sqrt{ \Phi''(x_{ \text{min} } )  \Phi''(x_{ \text{max} } ) } e^{-\Delta \Phi}, \label{eq:rate} \\
    \Phi(x) &:= \int^{x}_{0} \frac{F_{\text{eff}}(y)}{ D_{\text{eff}}(y) } dy. \label{eq:Phi_x}
\end{align}
\end{subequations}
Here \eqref{eq:rate} is the standard Kramer's expression for spatially dependent diffusion, where $x_{\rm{min}}$ is the position of the well minimum, while $x_{\rm{max}}$ is the position of the barier height maximum.
The key quantity in determining the rate $r$ is the quasi-potential $\Phi(x)$, since the exponential factor in Eq.~\eqref{eq:rate}  is completely governed by the effective barrier height $\Delta \Phi$.
$\Delta \Phi$ is reduced as $\tau$ increases (SM Sec.\ V), thereby enhancing the rate, mainly due to the first order $\lambda^{2} \tau$ contribution to the effective diffusion constant in Eq.~\ref{eq:Deff}.
Higher order corrections have an effect of increasing the barrier height $\Delta \Phi$ and decreasing the curvatures $\Phi''(x_{ \text{min} } ) $ and $\Phi''(x_{ \text{max} } )$, thereby causing the rate to be smaller than it would be if the $\tau$-dependence was resulting from the increased effective temperature alone. The resulting expressions are consistent with previous literature on the transition rates of the Active Ornstein-Uhlenbeck Particle \cite{LDT_transition}. 

\noindent {\it The large persistence regime: $ \frac{1}{k} \ll \tau $.}
In the limit of large $\tau$, the fast variable is the position $x$, such that $\mathcal{L}_{f} = \mathcal{L}_{x}$. The eigenvalue $l_{0}$ can be found by solving the equation $ \mathcal{L}_{x} | \rho_{0} \rangle = l_0(a) | \rho_{0} \rangle$ or $\mathcal{L}_{x}^{\dagger} | \phi_{0} \rangle = l_0(a) | \phi_{0} \rangle$ while keeping $a$ fixed. This will lead to $l_0 = - R_0(a)$, where $R_0(a)$ is an activity-dependent transition rate corresponding to the activity-dependent effective potential $ \varphi(x,a) = \int^{x}_{ x_{\rm{min}} } k F(x') + \lambda F_a(x',a) dx'$, with $a$ kept constant. Since we consider the AOUP case, we have $F_a(x,a) = \lambda a$. Approximations for $R_0(a)$, $| \phi_0 \rangle$ and $|\rho_0\rangle$ can be obtained using Kramer's theory in the rare event regime. We find a modified FP equation in $a$ of the form
\begin{equation}
    \tfrac{1}{\tau} \mathcal{L}_{ a, \rm{eff} } p_{\rm{a}}(a) - R(a) p_{\rm{a}}(a) = - r p_{\rm{a}}(a).
\label{eq:Eigenvalueproblem_activity}
\end{equation} 
Here, $ \mathcal{L}_{a,\rm{eff}} p_{\rm{a}} = \mathcal{L}_{a} p_{\rm{a}} + \partial_{a} ( 2 A p_{\rm{a}}) $, where $A$ is defined in Eq. \eqref{eq:activitydependentrate}. In the rare event limit ($r \ll 1$) we can approximate $\mathcal{L}_{a,\rm{eff}} \approx \mathcal{L}_{a}$ (SM Sec.\ IV). $p_{\rm{a}}(a) := \langle \phi_0 | P | p \rangle $ is the well-dependent density over the activities, and $R(a) = R_0(a) + \frac{1}{\tau} R_1(a) + \frac{1}{\tau^{2}} R_{2}(a) + O( \frac{1}{\tau^{3}})$ is the activity-dependent transition rate, i.e. the probability to transition given an activity $a$. Standard Kramer's theory results in
\begin{align}
    R_{0}(a) &= \tfrac{1}{2 \pi} \sqrt{ \varphi''(x_{\text{min}},a) \varphi''(x_\text{max},a) } e^{ - \Delta \varphi },
\label{eq:R0}
\end{align}
while $R_1(a)$ takes the form
\begin{subequations}
\begin{align}
    R_{1}(a) &= a A + B - 2 A',\\
    A(a) &:= \langle \phi_0 | \partial_{a} | \rho_0 \rangle, \quad B(a) := \langle \phi_0 | \partial_{aa} | \rho_0 \rangle.
\end{align}
\label{eq:R1}
\end{subequations}
The total rate $r$ is obtained by integrating both sides of Eq.\ \eqref{eq:Eigenvalueproblem_activity} over all $a$. Demanding that in the limits of $|a| \rightarrow \infty$ $p_{\rm{a}}(a) \rightarrow 0$ and that the current also vanishes, i.e. $J_{a} \rightarrow 0$, we find
\begin{align}
    r &= \int^{\infty}_{-\infty} R(a) p_{\rm{a}}(a) da. \label{eq:totalrate_largetau}
\end{align} 
The quantity $R(a)$ can be interpreted as the activity-dependent rate, i.e., the instantaneous probability for a particle with activity $a$ to cross the potential barrier per unit time. 

In the large-$\tau$ limit, as given by Eq.\ \eqref{eq:totalrate_largetau}, the only important quantity for calculating the rate $r$ is  $R(a) p_{\rm{a}}(a)$. Both $p_{\rm{a}}(a)$ and $R(a)$ are dependent on $\tau$, albeit in different ways. 
From Eq.\ \eqref{eq:Eigenvalueproblem_activity} we can identify two different dependencies on the parameter $\tau$. The activity-dependent rate $R(a)$ directly depends on $\tau$ as it is a series expansion in powers of $\frac{1}{\tau}$, 
directly affecting $r$ through Eq.\ \eqref{eq:totalrate_largetau}. However, for a given $R(a)$ the solution $p_{\rm{a}}(a)$ is also dependent on $\tau$ through the term $\frac{1}{\tau} \mathcal{L}_{a}$.
Consequently, two separate subregimes can be distinguished: one where the $\tau$-dependence of the rate $r$ can be approximated as the $\tau$-dependence of $R(a)$ only, and one where the $\tau$-dependence of the rate is resulting from the $\tau$-dependence of the density $p_{\rm{a}}(a)$ only. In between, there is a region of overlap where both $p_{ \rm{a} }(a)$ and $R(a)$ are approximately independent on $\tau$, which is where the rate is at its maximum. 
These two different regimes are the result of the separation of timescales in the rare event limit, where $r$ is typically orders of magnitude smaller than $1$. In this limit, $R(a)$ is also orders of magnitude smaller than unity. Since $\mathcal
L_{a} \sim \mathcal{O}(1)$, the terms $\frac{1}{\tau} \mathcal{L}_{a}$ and $R(a)$ become comparable only at very large $\tau$, leading to two effective sub-regimes within the large-$\tau$ regime. 
When both $R(a) \approx R_{0}(a)$, and $p_{ \rm{a} }(a) \approx \frac{1}{Z_a}e^{-v(a)}$, the maximum transition rate is $r_{\rm{max}} = \int^{\infty}_{-\infty} R_0(a) \frac{1}{Z_a}e^{-v(a)} da$, such that $r \le r_{ \rm{max} }$. Using this value, we can distinguish the two effective sub-regimes.

\noindent{\it The intermediate-to-large $\tau$ approximation: $ 1/k \ll \tau \ll 1/r_{\rm{max}} $. } 
When the persistence time is much smaller than the corresponding mean first passage time $1/r$ (i.e. $\tau \ll 1/r$) for a given $\lambda$, we make the approximation $\mathcal{L}_{a} p_{\rm{a}} \approx 0$ (SM Sec.\ IV). And since $r$ is bounded as $r \le r_{\rm{max}}$, the validity condition becomes $\tau \ll 1/r_{\rm{max}}$. Although the density $p_{\rm{a}}(a)$ is $\tau$-independent, $\frac{1}{\tau^{n}}$ corrections to $R(a)$ are still relevant, such that the only $\tau$-dependence of the rate $r = \int^{\infty}_{-\infty} \frac{1}{Z_{a}}e^{-v(a)} R(a) da$ is therefore in the activity-dependent rate $R(a)$. This subregime is denoted in Fig. \ref{fig:Conceptual_figure}, where one can see that the transition rate also tends to a maximum $r_{\rm{max}}$.

\noindent {\it The very large persistence time approximation: $\tau \gtrsim 1/r_{\text{max}} $.} 
Here, $\tau$ becomes comparable to the longest mean first passage time from a well, $\frac{1}{r_{\text{max}}}$. 
Then, higher-order corrections in $\frac{1}{\tau}$ to $R(a)$ become negligible, and $R(a) \approx R_{0}(a)$. 
In this regime, the $\frac{1}{\tau} \mathcal{L}_{a}$ term will not dominate over the $R(a)$ term in Eq.\ \eqref{eq:Eigenvalueproblem_activity}, so that the density $p_{\rm{a}}(a)$ will be $\tau$-dependent. In this regime, we have to take into account the influx of probability from neighbouring wells. The two-well version of Eq.\ \eqref{eq:Eigenvalueproblem_activity} (SM Sec.\ IV) is
\begin{align}
    \tfrac{1}{\tau} \mathcal{L}_{a} p_{\rm{a},\alpha} - \big( R_{ \alpha  \beta } + R_{ \beta \alpha } \big) p_{\rm{a}, \alpha} + R_{\beta \alpha} \tfrac{1}{Z}e^{ - v(a) } = 0.
    \label{eq:ODEVeryLargePersistence}
\end{align}

Here, $p_{\rm{a},\alpha}(a)$ is the density over activities while the particle is located in well $\alpha$. $R_{\alpha \beta}(a)$ is the activity-dependent rate from well $\alpha$ to well $\beta$. The total rate equals $r = \int^{\infty}_{-\infty} p_{\rm{a},\alpha}(a) R_{\alpha \beta}(a) da$. 
As $\tau$ is increased, the density $p_{\rm{a},\alpha}$ decreases in regions where the activity-dependent rate $R_{ \alpha \beta }(a)$ is high, thereby decreasing the overall product $p_{\rm{a},\alpha} R_{\alpha \beta}(a)$ and the total rate $r$.
The probability influx in well $\alpha$ for an activity value $a$ in a small time interval $\Delta t$ is obtained as $R_{\beta \alpha}(a) p_{\rm{a},\beta}(a) \Delta t$, where $p_{\rm{a},\beta}(a) = \frac{1}{Z}e^{-v(a)} - p_{\rm{a},\alpha}(a)$. Evolving this distribution in time using the time-dependent equation results in the First Passage Time Distribution (FPTD) (SM Sec.\ IV).

\begin{figure}[h!] 
\includegraphics[width=1.0\linewidth] {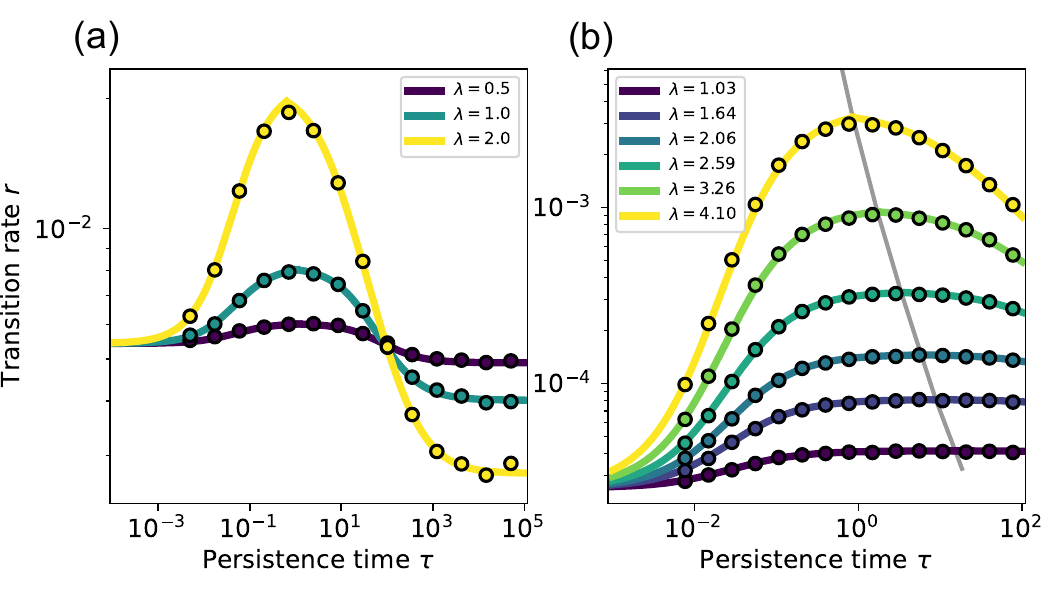} 
    \caption{ Comparison between analytically predicted transition rates (solid lines) and rates obtained by numerical simulation (markers) for barrier heights $k= 13$ (a) and $k=7$ (b), plotted as a function of persistence time $\tau$. The grey curve shows where the large-$\tau$ and very-large-$\tau$ approximations meet, and gives the maximum of the transition rate for given $\lambda$. The truly rare event case $k = 13$ shows increasing of the active rate compared to the passive rate as a function of $\tau$, up to a maximum. The less rare but more numerically attainable $k=7$ case shows that once $\tau$ becomes comparable to the passive transition rate, higher activity causes lower transition rates.} 
\label{fig:TransitionRates}
\end{figure}

\begin{figure}[h!]
    \centering
    \includegraphics[width=1\linewidth]{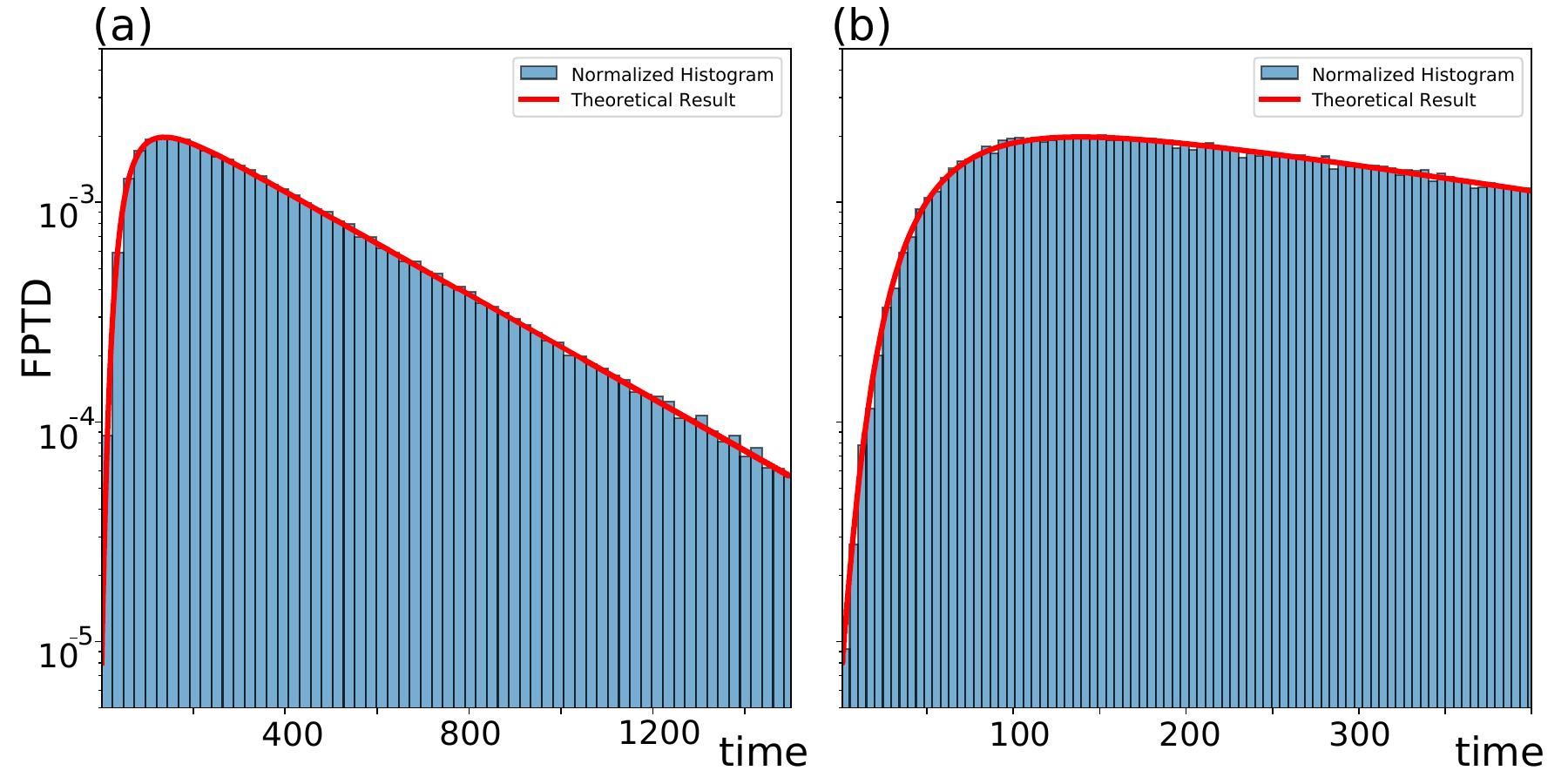}
    \caption{ First Passage Time Distribution (FPTD) for $k=13$, $\tau = 100$ and $\lambda = 0.5$. The FPTD is shown from $t=0$ to $t=1500$ (a) and from $t=0$ to $t=500$ (b). We can see that there is non-exponential behaviour in the FPTD until $t \sim \tau$. }
    \label{fig:FPTD}
\end{figure}

\noindent {\it Numerical simulations.} 
To validate our theoretical results, we compare them with transition rates obtained by numerically integrating the AOUP in a double-well potential. We use a second order stochastic Runge-Kutta scheme (SM Sec.\ VI) considering fixed barrier heights $k$ and perform  a parameter sweep  over $\lambda$ and $\tau$, with  $\tau$ spanning multiple orders of magnitude. Simulations were run until 20000 transitions between the wells were counted, from which the total rate is calculated.
In Fig.~\ref{fig:TransitionRates}(a) and (b), we present the transition rate  as a function of $\tau$ for several activity strengths $\lambda$ for intermediate ($k = 7$) and large ($k = 13$), respectively. We find excellent agreement across the full range of $\tau$, providing, to our knowledge, the first analytical prediction valid for arbitrary persistence time for thermal active particles. We also observe excellent agreement between the numerical data for the first-passage-time distribution (FPTD) and the theoretically derived result in Fig.~\ref{fig:FPTD}. The figure illustrates that the FPTD becomes exponential only on timescales $t \gg \tau$.

\noindent {\it Conclusion and outlook.}
In this work, we have developed a projection-operator formalism to compute the transition rates for a generalized class of active particles in both the low and large persistence-time limits including  thermal fluctuations. In the small persistence-time limit, we obtain an effective Fokker-Planck equation in the position $x$ alone, with an effective diffusion $D_{\rm{eff}}(x)$ and, at higher order, an effective drift $F_{\rm{eff}}(x)$. This equation  can be treated using standard Kramer's theory, yielding the rate as an expansion in powers of $\tau$. In the large persistence-time limit, however, we obtain a modified Fokker-Planck in the activity $a$ acting on the well dependent density $p_{a}(a)$, which includes a sink term $R(a)$ that we interpret as the probability to transition given an activity $a$. We find that the rate $r$ is completely controlled by the quantity $ p_{a}(a) R(a)$, and identify two sub-regimes. These findings are summarized in Fig. \ref{fig:Conceptual_figure}.
In the intermediate-to-large persistence-time-regime, $p_{\rm{a}}(a)$ is given by the global steady state probability density of the activity, up to normalization factor, such that $p_{\rm{a}}(a) \propto \int_{ -\infty }^{\infty} p(x,a) dx \propto e^{-v(a)}$. The $\tau$-dependence results instead from the correction terms to $R(a)$.
In the very-large persistence-time-regime, the $O(1/\tau)$ corrections to $R(a)$ become negligible. Instead, the well-dependent density $p_{\rm{a}}(a)$ becomes biased, and does not equal the global density, which results in the $\tau$-dependence of the rate. In this regime, FPTDs are exponential only on timescales larger than $\tau$.
Our extension of Kramers' theory for active systems provides a framework to calculate transition rates for higher dimensional systems, if one is able to find appropriate reaction coordinate resulting in a lower dimensional equation of motion. This could be applied to describe dynamical phase-transitions such as the Motility Induced Phase Separation phenomena, or the collapse of active polymers.
Furthermore, the projection framework may be extendable to cases in which the driving variable also depends on the driven variable, thereby broadening its applicability to a wider class of multi-stable driven systems with stochastic transitions \cite{Gene_switching_paper, Cortical_states_paper}.

\bibliography{bibliography}

\onecolumngrid 





\section{Non-dimensionalization}

We start with the dimensional version of the generalized active particle Langevin equations:
\begin{align}
    \frac{d x^{*}}{dt^{*}} &= - \frac{k^{*}}{\gamma} F(x^{*}) - \frac{\lambda^{*}}{\gamma} F_{a}(x^{*},a)
    + \sqrt{\frac{2 k_{B} T}{\gamma}} \, \eta_{\rm{x}}(t^{*}), \\
    \frac{d a}{dt^{*}} &= - \frac{1}{\tau^{*}} v'(a) + \sqrt{\frac{2}{\tau^{*}}} \, \eta_{a}(t^{*}),
\end{align}
where $\langle \eta_{\rm{x}}(t_1^{*}) \eta_{\rm{x}}(t_2^{*}) \rangle = \delta(t_1^{*} - t_2^{*})$ and $\langle \eta_{\rm{a}}(t_1^{*}) \eta_{\rm{a}}(t_2^{*}) \rangle = \delta(t_1^{*} - t_2^{*})$. We have chosen $a$ to be dimensionless, such that the non-dimensionalization will not affect it.
The corresponding Fokker-Planck equation takes the form
\begin{align}
    \partial_{t^{*}} p^{*} &=
    \partial_{x^{*}} \!\left[
        \left(
            \frac{k^{*}}{\gamma} F(x^{*}) + \frac{\lambda^{*}}{\gamma} F_{a}(x^{*},a)
        \right) p^{*}
    \right]
    + \frac{k_{B} T}{\gamma} \, \partial_{x^{*}}^{2} p^{*} 
    + \frac{1}{\tau^{*}} \partial_{a} \!\left[ v'(a) p^{*} \right]
    + \frac{1}{\tau^{*}} \partial_{a}^{2} p^{*}.
\end{align}
Here, $p^{*}$ denotes the dimensionful density.
To render the equations dimensionless, we choose the characteristic length $L$ (the distance between the well and origin), the thermal energy $k_{B}T$ as the energy scale and characteristic time scale as $\frac{\gamma L^{2}}{k_{B} T}$.
 We rescale the variables as
\begin{align}
    x^{*} &= L x, &
    t^{*} &= \frac{\gamma L^{2}}{k_{B} T} \, t, &
    \tau^{*} &= \frac{\gamma L^{2}}{k_{B} T} \, \tau, \\
    k^{*} &= \frac{k_{B} T}{L} \, k, &
    \lambda^{*} &= \frac{k_{B} T}{L} \, \lambda, &
    p^{*}(x^{*},a,t^{*}) &= \frac{1}{L} p(x,a,t),
\end{align}
where the corresponding derivatives take the form
\begin{align}
    \partial_{t^{*}} &= \frac{k_{B} T}{\gamma L^{2}} \, \partial_{t}, &
    \partial_{x^{*}} &= \frac{1}{L} \, \partial_{x}, & 
    \partial_{x^{*}}^{2} &= \frac{1}{L^{2}} \, \partial_{x}^{2}. 
\end{align}
Dividing both sides of the equation by the factor $ \frac{k_B T}{ \gamma L^{3} } $, we obtain 
\begin{align}
    \partial_{t} p &=
    \partial_{x} \!\left[
        \left(
            k F(x) + \lambda F_{a}(x,a)
        \right) p
    \right]
    + \partial_{x}^{2} p
    + \frac{1}{\tau} \partial_{a} \!\left[ v'(a) p \right]
    + \frac{1}{\tau} \partial_{a}^{2} p .
\end{align}
Thus, we have derived the Fokker-Planck equation corresponding to Eqs. (1a) \& (1b) in the main text.

\section{Projection formalism and application of the Feshbach-Schur map}

We start with a generic Fokker-Planck operator where there is a fast and slow part
\begin{align}
    \mathcal{L} = \mathcal{L}_{f} + \mathcal{L}_{s}.
\end{align}
This can be made explicit by introducing a small parameter $\epsilon \ll 1$, such that $\mathcal{L}_{s} = \epsilon \mathcal{L}_{s}'$, where $ \mathcal{L}_{f} $ is of the order of $ \mathcal{L}_{s}' $. In this case it follows that the operator $\mathcal{L}_{f}$ will result in faster dynamics than the operator $\epsilon \mathcal{L}_{s}$. We denote the left principal eigenstate of $\mathcal{L}_f$ as $|\rho_0 \rangle$ and the right eigenstate as $| \phi_0 \rangle$, such that
\begin{align}
    \mathcal{L}_{f} | \rho_{0} \rangle &= l_{0} | \rho_{0} \rangle, \\
    \mathcal{L}_{f}^{\dagger} | \phi_{0} \rangle &= l_{0} | \phi_{0} \rangle,
\end{align} 
where $\mathcal{L}_{f}^{\dagger}$ is the backward Fokker-Planck operator, and the adjoint of $\mathcal{L}_{f}$. The left and right eigenfunctions satisfy the bi-orthogonality condition $\langle \phi_{0}| \rho_{0} \rangle=  1$. $l_0$ is the highest eigenvalue. Since all eigenvalues are negative, it has the smallest absolute value. The projection operator is defined as and equal to
\begin{align}
    P := \lim_{t \rightarrow \infty} e^{t(\mathcal{L}_{f} + l_{0})} = | \rho_{0} \rangle \langle \phi_{0}|.
\end{align} 
Applying $P$ to any of the higher modes $| \rho_n \rangle$, where $n \ge 1$, will result in $P | \rho_n \rangle = 0$. Therefore, $P$ extracts only the component of the slowest eigenmode.

We start with our Fokker-Planck operator of interest. We will consider in the next sections the small $\tau$ regime and the large $\tau$ regime. In the small $\tau$ regime, $a$ is the fast variable and $x$ is the slow variable, $\epsilon = \tau$ and $\mathcal{L}_{f} = \mathcal{L}_{a}$ while $\mathcal{L}_{s} = \mathcal{L}_{x}$. In the large $\tau$ regime, $x$ is the fast variable and $a$ is the slow variable, $\epsilon = 1/\tau$ and $\mathcal{L}_{f} = \mathcal{L}_{x}$ while $\mathcal{L}_{s} = \mathcal{L}_{a}$.
Near the stable points of a well, the relaxation rate of $\mathcal{L}_{x}$ is given by the prefactor $w$ of the (locally) harmonic potential $w (x-x_{\rm{min}})^2 /2$, which in our convention scales with $k$. The relaxation time $1/w$ then scales as $1/k$. Therefore, the small persistence time limit is characterized by the condition $\tau \ll 1/k $, such that the small parameter is given by $ k \tau \ll 1$. Conversely, the large persistence time limit is characterized by the condition $1/k \ll \tau $, such that the small parameter is given by $ 1/(k \tau) \ll 1$. To make the expansions less cluttered however, we will simply write $ \tau $ and $ 1 / \tau $ respectively for the small parameters, where factors if $k$ will be absorbed into the corresponding coefficients. We should notice however, that this choice does not change the fact that the crossover still occurs at around $k \tau \sim 1$.

We are interested in the transition rate in the rare event limit, where there is a large barrier in $x$. To determine this rate, we will pick an absorption point $x_{c}$ that is placed far enough over the barrier that the recrossing probability is negligible. We have 
\begin{align} 
    p(x_{c},a) &= 0. 
\label{eq:absorbingboundarycondition}
\end{align}
Particularly, we are looking for the principal eigenvalue of the full Fokker-Planck operator, which will give the rate at which the probability leaves a well in the quasi-steady state. Therefore, we are looking for the highest value $\Lambda = \Lambda_0$ that solves the equation
\begin{align}
    (\mathcal{L} - \Lambda) p &= 0. 
\label{eq:GroundStateTotal}
\end{align}
We will aim to do this using projections in the fast operator $\mathcal{L}_{x}$, which acts on the fast variable.

We will proceed by following the steps of what is called the Feshbach-Schur map in physics \cite{Feshbach}.
Using the definition of the orthogonal projector $Q := I - P$, we have $ | p \rangle = I |p \rangle = (P + Q) | p \rangle $. Applying the operators $P$ and $Q$ to both sides of Eq. \ref{eq:GroundStateTotal}, we obtain
\begin{align}
    P (\mathcal{L} - \Lambda_0) (P + Q) |p \rangle &= 0, \\
    Q (\mathcal{L} - \Lambda_0) (P + Q) |p \rangle &= 0.
\end{align}
We proceed by distributing the operators and clearing up the notation by writing the projection of $|p\rangle$ as $|u \rangle := P |p \rangle$, while we write the remainder as $| v \rangle := Q | p \rangle$. We define the operators $A$, $B$, $C$ and $D$ as:
\begin{align}
    \underbrace{P (\mathcal{L} - \Lambda_0) P}_{=:A} \underbrace{ ( P | p \rangle ) }_{| u \rangle} + \underbrace{ P (\mathcal{L} - \Lambda_0 ) Q}_{=:B} \underbrace{( Q | p \rangle )}_{ | v \rangle } &= 0 \label{eq:P_equation} \\
    \underbrace{Q (\mathcal{L} - \Lambda_0 ) P}_{=:C} \underbrace{ (P | p \rangle)  }_{ | u \rangle } + \underbrace{ Q (\mathcal{L} - \Lambda_0) Q}_{=:D} \underbrace{ (Q | p \rangle) }_{| v \rangle} &= 0 \label{eq:Q_equation}
\end{align}
We then express our quantity $ | v \rangle $ in terms of $ | u \rangle $ by applying the inverse of operator $D$ to both sides of equation \ref{eq:Q_equation}. We obtain the equation $| v \rangle = D^{-1} C | u \rangle$. We can make further simplifications using identities of the projection and orthogonal projection operators. The relevant identities are $PQ = QP = 0$, $P \mathcal{L}_{f} = \mathcal{L}_f P = l_0 P $ and $P^2 = P$. 

Using $P Q = Q P = 0$, it follows that $P(\mathcal{L}-\Lambda_0) Q = P \mathcal{L} Q$ and $Q(\mathcal{L}-\Lambda_0) P = Q \mathcal{L} P$. Furthermore, by making use of the identities $P \mathcal{L}_{f} Q = Q \mathcal{L}_{f} P = 0$, we obtain $P \mathcal{L} Q = P \mathcal{L}_{s} Q $. 
Inserting this into equation \ref{eq:P_equation} and writing out the operators fully, we obtain 
\begin{align}
    \langle \phi_{0} | P  \big( \mathcal{L} - \Lambda_0 \big) P | p \rangle - \langle \phi_{0} | P \mathcal{L}_{s} Q \big( Q ( \mathcal{L} - \Lambda_0) Q \big)^{-1} Q \mathcal{L}_{s} P | p \rangle = 0.
\end{align} 
Here, we also applied $\langle \phi_0 |$ to both sides to fully integrate out the fast variable, and made use of the identity $P^2 = P$. We will make use of the fact that $\langle \phi_0 | P = \langle \phi_0 | \rho_0 \rangle \langle \phi_0 | = \langle \phi_0 |$ to simplify $\langle \phi_{0} | P  \big( \mathcal{L} - \Lambda_0 \big) P | p \rangle = \langle \phi_{0} | \big( \mathcal{L} - \Lambda_0 \big) P | p \rangle$. Since the rate is the negative of the principal eigenvalue, we will henceforth write $\Lambda_0 = -r$.

\section{Small persistence time regime} \label{LowPersistenceTime}
In the small persistence time regime, the fast operator is $\mathcal{L}_{a}$, while $\tau$ is the small parameter. The projection operator $P$ takes the form $P = | \rho_0 \rangle \langle \phi_0 |$, in absence of an absorbing boundary in $a$, we have $| \phi_0 \rangle = 1$ and $| \rho_0 \rangle = \frac{1}{Z_{a}} e^{-v(a)}$. The operation $\langle f | \cdot$ is now shorthand notation for the integration $\int^{\infty}_{-\infty} f \cdot da$. We define the quantity $\langle \phi_0|p\rangle$ as
\begin{align}
    p_{\rm{x}}(x) := \langle \phi_0 | p \rangle = \int^{\infty}_{-\infty} p(x,a) da.
\end{align}
The projected density takes the form
\begin{align}
    P | p \rangle = |\rho_0\rangle \langle \phi_0| p \rangle = | \rho_0 \rangle p_{\rm{x}}(x),
\end{align}
from which it follows that $\langle \phi_0 | P | p \rangle = p_{ \rm{x} }(x)$. 

We now turn to the eigenvalue equation. Since the lowest eigenvalue for $\mathcal{L}_{\rm{a}}$ equals zero, we have $P \mathcal{L}_{a} = \mathcal{L}_{a} P = 0 $, which we use to simplify $ P \mathcal{L} P = P \left( \mathcal{L}_{x} + \frac{1}{\tau} \mathcal{L}_{a} \right) P = P \mathcal{L}_x P $, and obtain
\begin{align}
    \langle \phi_{0} | P \mathcal{L}_{x} P | p \rangle - \langle \phi_{0} |\left( P \mathcal{L}_{x} Q \left( \tfrac{1}{\tau} \mathcal{L}_{a} + Q (\mathcal{L}_{x} + r ) Q \right)^{-1} Q \mathcal{L}_{x} P \right) P |p \rangle = - \langle \phi_0 | r P |p \rangle. 
\label{eq:lowpersistenceeigenvalue}
\end{align} 
We rewrite $\tfrac{1}{\tau} \mathcal{L}_{a} + Q (\mathcal{L}_{x} + r ) Q = \frac{1}{\tau} \mathcal{L}_{a} \big(I + \tau \mathcal{L}_{a}^{-1} Q (\mathcal{L}_{x} + r ) Q \big) $, where $\mathcal{L}_{a}^{-1}$ is defined as the psuedo-inverse of $\mathcal{L}_{a}$ through the relation $\mathcal{L}_{a}^{-1} \mathcal{L}_{a} = Q$. We make use of the relation $(AB)^{-1} = B^{-1} A^{-1}$, and perform a standard Neumann series in the small parameter $\tau$ to obtain
\begin{align}
    \left( \frac{1}{\tau} \mathcal{L}_{a} + Q (\mathcal{L}_{x} + r ) Q \right)^{-1} &= \big(I + \tau \mathcal{L}_{a}^{-1} Q (\mathcal{L}_{x} + r ) Q \big)^{-1} \tau \mathcal{L}_{a}^{-1} \\
    &= \tau \mathcal{L}_{a}^{-1} - \tau^{2} \mathcal{L}_{a}^{-1} Q (\mathcal{L}_{x} + r ) Q \mathcal{L}_{a}^{-1} + O \left( \tau^{3} \right).
\end{align}
Inserting this into Eq. \eqref{eq:lowpersistenceeigenvalue} and making use of the relation $P | p \rangle = | \rho_0 \rangle p_{ \rm{x} }$, results in the expanded equation
\begin{align}
    \mathcal{L}_{x,0} p_{\rm{x}} + \tau \mathcal{L}_{x,1} p_{\rm{x}} + \tau^{2} \mathcal{L}_{x,0} p_{\rm{x}} + O(\tau^{3}) = -r p_{\rm{x}},
\end{align}
where $\mathcal{L}_{x,n}$ take the form
\begin{align}
    \mathcal{L}_{x,0} &= \langle \phi_0 | \mathcal{L}_{x} | \rho_0 \rangle, \\ 
    \mathcal{L}_{x,1} &= - \langle \phi_0 | (P \mathcal{L}_{x} Q) \mathcal{L}_{a}^{-1} (Q \mathcal{L}_{x} P) | \rho_0 \rangle, \\ 
    \mathcal{L}_{x,2} &= \langle \phi_0 | (P \mathcal{L}_{x} Q) \mathcal{L}_{a}^{-1} Q (\mathcal{L}_{x} + r ) Q \mathcal{L}_{a}^{-1} (Q \mathcal{L}_{x} P) | \rho_0 \rangle.
\end{align}
We now turn to the special case of the Active Ornstein Uhlenbeck particle, where we define the passive operator
\begin{align}
    \mathcal{L}_{x}^{ \rm{eq} } \cdot = \partial_{x} (- F(x) \cdot) + \partial_{xx} ( \cdot ),
\end{align}
which has no dependencies on the variable $a$, and also doesn't act on $a$. The FP-operator acting on $x$ becomes
\begin{align}
    \mathcal{L}_{x} \cdot = \mathcal{L}_{x}^{ \rm{eq} } \cdot + \lambda a \partial_{x} \cdot.
\end{align}
We now turn to the leading order effective operator $\mathcal{L}_{x,0}$
\begin{align}
    \mathcal{L}_{x,0} = \langle \phi_0 | \mathcal{L}_{x} | \rho_0 \rangle = \langle \phi_0 | \rho_0 \rangle \mathcal{L}_{x}^{\rm{eq}} + \langle \phi_0 | a |\rho_0 \rangle \lambda \partial_{x},
\end{align}
where $\langle \phi_0 | a |\rho_0 \rangle = \int^{\infty}_{-\infty} \frac{1}{Z_{a}}e^{ - a^{2}/2 } a \,da = 0$, while $\langle \phi_0 | \rho_0 \rangle = 1$, such that we have to $0$-th order the passive FP equation
\begin{align}
    \mathcal{L}_{x,0} = \partial_{x} (-F \cdot ) + \partial_{xx}.
\end{align}
To compute the higher order operator corrections, we compute the following simplifications. 
\begin{align}
    P \mathcal{L}_{x} Q | f \rangle &= - \lambda \partial_{x} P a P | f \rangle, \\
    Q \mathcal{L}_{x} P | f \rangle &= - \lambda a \partial_{x} P | f \rangle ,
\end{align}
which stems from the fact that $\mathcal{L}_{x}^{\rm{eq}}$ commutes with $P$ and $Q$, and the fact that $PQ = QP = 0$, such that only the $\lambda a \partial_{x}$ part of $\mathcal{L}_{x}$ survives. We then obtain for the 1st and 2nd order operator corrections
\begin{align}
    \mathcal{L}_{x,1} |p\rangle &= - \langle \phi_0 |\lambda^{2} \partial_{x} P a \mathcal{L}_{a}^{-1} a \partial_{x} P |p\rangle , \\
    \mathcal{L}_{x,2} |p \rangle &= \langle \phi_0 | \lambda^{2} \partial_{x} P a \mathcal{L}_{a}^{-1} Q (\mathcal{L}_{x} + r ) Q \mathcal{L}_{a}^{-1} a ( \partial_{x} P |p\rangle ).
\end{align} 
The first order correction becomes
\begin{align}
    \mathcal{L}_{x,1} | p \rangle = -\lambda^{2} \partial_{xx} \langle \phi_0 | P a \mathcal{L}_{a}^{-1} a P | p \rangle = - \lambda^{2} \partial_{xx} p_{\rm{x}} \langle \phi_0 | P a \mathcal{L}_{a}^{-1} a | \rho_0 \rangle = \lambda^{2} \partial_{xx} p_{ \rm{x} },
\end{align} 
where we made use of the fact that $a \rho_0 $ is the first eigenfunction of the OU operator $\mathcal{L}_{a}$ with eigenvalue $-1$, such that $\mathcal{L}_{a}^{-1} a | \rho_0 \rangle = \frac{1}{-1} a | \rho_0 \rangle = - a | \rho_{0} \rangle$, while $\langle \phi_0 | -a^{2} | \rho_{0} \rangle = -1$. This term is an effective contribution to the temperature the particle is subjected to. The second order operator correction becomes
\begin{align}
    \mathcal{L}_{x,2} p = \left( \lambda^{2} \partial_{x} (\mathcal{L}_{x} - \Lambda) \partial_{x} p_{ \rm{x} } \right) \langle \phi_0 | P a \mathcal{L}_{a}^{-1} \mathcal{L}_{a}^{-1} a | \rho_{0} \rangle,
\end{align}
where the term proportional to $-\lambda^{3} \partial_{x}^{3} p_{\rm{x} }$ involves integrating over an uneven polynomial in $a$ multiplied by $e^{-a^2/2}$, which equals $0$. We furthermore make use of the commutation relation between $\mathcal{L}_{x}^{\rm{eq}}$ and $\partial_x$, resulting in $(\mathcal{L}^{\rm{eq} }_{x} + r) \partial_{x} p_{\rm{x}} = \partial_{x} (\mathcal{L}^{\rm{eq} }_{x} + r) p_{\rm{x}} + \partial_{x}(F' p_{\rm{x}})$. Inserting the leading order relation $(\mathcal{L}_{x}^{\rm{eq}} + r) p_{ \rm{x} } = 0 + O(\tau)$, we obtain
\begin{align} 
    \mathcal{L}_{x,2} p = \lambda^{2} \partial_{xx}(F'p_{\rm{x}}) + O(\tau).
\end{align} 
Multiplying with $\tau^{2}$ means that the $O(\tau)$ correction in $\mathcal{L}_{x,2} p$ results in an $O(\tau^{3})$ term that we neglect in our analysis, since we only proceed until second order. 
We finally obtain
\begin{align}
    \partial_{x} (- F(x) p_{\rm{x}}) + \partial_{xx} p_{\rm{x}} + \tau \lambda^{2} \partial_{xx}p_{\rm{x}} + \tau^{2} \lambda^{2} \partial_{xx}(F'p_{\rm{x}}) = -r p_{ \rm{x} },
\end{align}
thus leading to Eqs. \eqref{eq:eigenvalueproblem_xspace} and \eqref{eq:Deff}. We note that these expressions are equivalent to those found by the time-dependent projection formalism found in the case where there is no passive diffusion ($k_{B} T = 0$) \cite{ProjectionOperatorFasta}, and also the same expression as the one found by Fox \cite{Fox_paths}. In these papers, the higher order terms in $\tau$ were re-summed for the branch involing only terms that are quadratic in the operator $\lambda a \partial_{x}$. We do not perform this procedure since we are interested in polynomials up to second order in $\tau$ only for the two-point Pade approximation, since the resummed expression would need to be Taylor expanded up to second order in $\tau$ in any case.

\section{Large persistence time limit} \label{largepersistence time} 
In the limit of large persistence time the small parameter takes the form $\frac{1}{\tau} \ll 1$. Now, $\mathcal{L}_{x}$ is the fast operator while $\mathcal{L}_{a}$ is the slow operator. Given the boundary condition in Eq. \ref{eq:absorbingboundarycondition}, the negative of the principal eigenvalue solution to $\mathcal{L}_{f} | \rho_{0} \rangle = l_0 | \rho_0 \rangle$ will give the activity dependent transition rate $R_{0}(a)$ in the limit of $\tau \rightarrow \infty$, and gives the zeroth order contribution to the activity dependent rate $R(a)$. The full projected equation becomes 
\begin{align}
    \langle \phi_0 | \mathcal{L} P | p \rangle + \tfrac{1}{\tau^{2}} 
    \langle \phi_0 | P \mathcal{L}_{a} Q \big( Q ( \mathcal{L} + r) Q \big)^{-1} Q \mathcal{L}_{a} P | p \rangle = - r \langle \phi_0 | P | p \rangle 
\label{eq:FullSlowActivityProjected}
\end{align}
We will denote the projected density as 
\begin{align}
    P |p \rangle &= |\rho_{x,0} \rangle \langle \phi_{x,0} | p \rangle  = f(a) | \rho_{x,0} \rangle, \\
    f(a) &:= \langle \phi_{x,0} | p \rangle = \int^{ x_c }_{-\infty} \phi_0(x,a) p(x,a) dx 
\end{align}
We proceed by expanding Eq. \ref{eq:FullSlowActivityProjected} in powers of $\frac{1}{\tau}$, and cutting off the expansion at an order of choice. We obtain
\begin{align}
    \langle \phi_0 | \mathcal{L}_{x} f(a) | \rho_0 \rangle = f(a) l_0(a) \langle \phi_0 | \rho_0 \rangle = f(a) l_0(a)
\end{align} 
To order $\tfrac{1}{\tau}$, the equation for the principal eigenvalue then becomes
\begin{align}
    l_{0}(a) f(a) + \tfrac{1}{\tau} \langle \phi_{0} | \mathcal{L}_{a} \big( f(a) | \rho_{0} \rangle \big) = \Lambda_{0} f(a). 
\label{eq:unsplitted_rate_eff} 
\end{align}
We have the operator $\mathcal{L}_{a}$ operating on a product of functions, which can be split as 
\begin{align}
    \langle \phi_{x,0}^{c} | \mathcal{L}_{a} ( f(a) \rho_{x,0}^{c} \rangle ) &= \mathcal{L}_{a} f(a) + f(a) \langle \phi_{x,0}^{c} | (v' \partial_{a} + \partial_{aa} ) | \rho_{x,0}^{c} \rangle + 2 \langle \phi_{x,0}^{c} | \partial_{a} | \rho_{x,0}^{c} \rangle \partial_{a} f(a).
\end{align} 
We define
\begin{align}
    A(a) &:= \langle \phi_0 | \partial_{a} | \rho_{0} \rangle = \int^{x_c}_{- \infty} \phi_0(x,a) \partial_{a} \rho_0(x,a) dx, \\
    B(a) &:= \langle \phi_0 | \partial_{aa} | \rho_{0} \rangle = \int^{x_c}_{- \infty} \phi_0(x,a) \partial_{aa} \rho_0(x,a) dx.
\end{align}
Inserting this into Eq. \ref{eq:unsplitted_rate_eff} and writing $l_{0} = -R_{0}, \Lambda_{0} = - r$, we obtain the equation:
\begin{align} 
    - R_{0}(a) f(a) + \tfrac{1}{\tau} \mathcal{L}_{a} f(a) + \tfrac{1}{\tau} (v' A + B) f(a) + \tfrac{1}{\tau} 2 A \partial_{a} f(a) = -r f(a).
\label{eq:FirstOrderFinal} 
\end{align} 
We write $A f'$ as $A f' = (Af)' - A'f$ using the product rule, and obtain an effective Fokker-Planck equation with a sink term
\begin{align}
    \frac{1}{\tau} \underbrace{ \left( \partial_a \left( [v' + 2A] f  \right) + \partial_{aa} f \right)}_{ \mathcal{L}_{a,\text{eff}} := } - R_{0}(a) f(a) + \frac{1}{\tau} \underbrace{ \left( v' A + B - 2 A' \right) }_{ - R_{1}(a) := } f(a) = - r f(a).
    \label{eq:FirstOrderActivityFull}
\end{align} 
Here, we have $ \mathcal{O} \big( R_{1}(a) \big) \sim \mathcal{O} \big( R_{0}(a) \big) $. \\
When one knows the solution $f(a)$, the rate for any equation of the type $\frac{1}{\tau} \mathcal{L}_{a,\text{eff}} f(a) - R(a) f(a) = -r f(a)$ can be found by integrating over all activities $a$. We have zero flux conditions in the limits $|a| \rightarrow \infty$, such that we obtain
\begin{align} 
    r = \int^{\infty}_{-\infty} R(a) f(a) da 
\end{align}
as noted in Eq. \eqref{eq:totalrate_largetau} in the main text.

\subsection{Intermediate-to-large persistence time limit}

Since we are in the rare event limit, we have typically have that $R(a)$ is small. On the other hand, the operator $\mathcal{L}_{a,\text{eff}}$ contains the term $\partial_{a} ( a\cdot) = v'(a) \cdot + v(a) \partial_{a}(\cdot)$, where the most relevant activities $a$ are at order $1$, such that this the operator $\mathcal{L}_{a,\text{eff}}$ is of order $1$. In fact, for the AOUP we have exactly $v'(a) = 1$. This then implies that $\frac{1}{\tau} \mathcal{L}_{a,\text{eff}}$ is of order $\frac{1}{\tau}$. Assuming small enough $\tau$ such that typically $R(a) \ll \frac{1}{\tau}$, this implies $R(a) f(a) \ll \frac{1}{\tau} \mathcal{L}_{a} f(a)$. Since $r = \int^{\infty}_{-\infty} f(a) R(a) da$, this already implies that $r f(a) \ll \mathcal{L}_{a,\text{eff}} f(a) $. Therefore, our equation in the activity becomes $\mathcal{L}_{a,\text{eff}} f(a) \approx 0$. We thus have the solution $f(a) \approx \frac{1}{Z_{a}} e^{-v(a) - 2 \int A da } $. In this limit, $A \sim \mathcal{O}(R_{0}) \ll 1$, such that we can approximate $f(a) \approx \frac{1}{Z_{a}} e^{-v(a)} (1-2\int A da) $ 
\begin{align} 
    r &\approx \int^{\infty}_{-\infty} \tfrac{1}{Z_{a} } e^{ - v(a) } \left( 1 - 2 \int A da \right) \left( R_{0} + \tfrac{1}{\tau} R_{1} \right) da \\
    &\approx \int^{\infty}_{-\infty} \tfrac{1}{Z_{a} } e^{ - v(a) }  \left[ R_{0}(a) + \tfrac{1}{\tau} \big( -v'(a) A(a) - B(a) + 2 A' \big)  \right] da.
\end{align}
Here we neglected the $\int A da \sim \mathcal{O}(R_0)$ term, because it only results in $\mathcal{O}(R_0^{2})$ terms. We can write the $A'$ term in terms of $A$ through integration by parts
\begin{align}
    \int^{\infty}_{-\infty} 2 A' e^{-v(a)} da = \underbrace{ 2 A e^{-v(a)} \big|^{\infty}_{-\infty} }_{= 0} - \int^{\infty}_{-\infty} 2 A(a) [-v'(a) e^{-v(a)}] da, 
\end{align}
such that we finally obtain
\begin{align}
    r \approx \int^{\infty}_{-\infty} \tfrac{1}{Z_{a} } e^{ - v(a) }  \left[ R_{0}(a) + \tfrac{1}{\tau} \big( v'(a) A(a) - B(a) \big) \right] da,
\label{eq:FinalSimplification}
\end{align} 
as mentioned in the main text in Eqs. \eqref{eq:R0} and \eqref{eq:R1}.
We also could have obtained Eq. \eqref{eq:FinalSimplification} by simply holding that $ \tfrac{1}{\tau}\mathcal{L}_{a} f(a) \approx 0$ in the rare event limit, finding that $f(a) \approx \frac{1}{Z_{a}} e^{-v(a)}$ and inserting this into Eq. \eqref{eq:FirstOrderFinal}. In general, $f(a)$ will equal $p_{\rm{a}}(a)$ plus corrections of order $r$, such that in the rare event regime we can approximate $f(a) = p_{\rm{a}}(a)$ in all cases. 
We now obtain a rate of the form 
\begin{align}
    r = b_0 + \frac{b_1}{\tau}.
    \label{eq:rate_nonexp_largetau}
\end{align}
For the Ornstein-Uhlenbeck Particle we have $v'(a) = a$, such that $b_0$ and $b_1$ are explicitly given by
\begin{align}
    b_0 &= \int^{\infty}_{-\infty} \frac{ e^{-a^2} }{\sqrt{2 \pi} } R_0(a) da, \\
    b_1 &= \int^{\infty}_{-\infty} \frac{ e^{-a^2} }{\sqrt{2 \pi} } [a A(a) - B(a)] da.
\end{align}
If the active force $\lambda a$ is smaller than $k$ such that $k - \lambda a \gg 1$, we can use Kramer's theory to find $R_0(a) = \sqrt{V''(x_{ \rm{min} } ) V''(x_{ \rm{max} })} e^{-\Delta \varphi(a)}$, where $\varphi = V(x) + \lambda a x$, and $\Delta \varphi(a)$ gives the difference between its minimum and maximum values. Note that $\varphi'' = V'' + (\lambda a x)'' = V''$. One can furthermore perform boundary layer calculations to find analytic approximations for $a_1$ under these conditions. To test the effectiveness of the projection-operator formalism we shall calculate these quantities numerically instead. 

\subsection{ Very large persistence time regime } \label{VeryHighPersistence}
In the very large persistence time regime, where $\tau$ is comparable to $1/r$, we can not approximate $p_{\rm{a}}$ as the solution to $\mathcal{L}_{a} p_{\rm{a}}(a) = 0$. Instead, it is important to take into account the biasing of activity probability density $p_{\rm{a}}$. This is a result of both outgoing and incoming probability. Therefore, if there are multiple metastable points in the potential under consideration, one must couple these probability densities to each other. In the special case where there are two metastable points in the potential, as in our double well potential, we will denote the well at $x=-1$ as $\alpha$, while we denote the well at $x=1$ as $\beta$. Then, we denote the activity probability densities in each well as $p_{\rm{a},\alpha}(a)$ and $p_{\rm{a},\beta}(a)$ respectively. For each density, we will follow the same procedure but now we add the loss term in well $\alpha$ as a gain term in well $\beta$, and vice versa. The corresponding equations of motion are
\begin{align}
    \tfrac{1}{\tau} \mathcal{L}_{a} p_{\rm{a},\alpha} - R_{\alpha \beta}(a) p_{\rm{a},\alpha} + R_{\beta \alpha}(a) p_{\rm{a},\beta} = 0, \\
    \tfrac{1}{\tau} \mathcal{L}_{a} p_{\rm{a},\beta} - R_{\beta \alpha}(a) p_{\rm{a},\beta} + R_{\alpha \beta}(a) p_{\rm{a},\alpha} = 0.
\end{align} 
For the example of a double well potential, where the barrier top is located at $x=0$, we have $p_{ \rm{a},\alpha }(a) = \int^{0}_{-\infty} p(x,a) dx$ and $p_{ \rm{a},\beta }(a) = \int^{ \infty }_{ 0 } p(x,a) dx$.
For the combined density $p_{ \rm{a} }(a) = p_{\rm{a},\alpha}(a) + p_{\rm{a},\beta}(a)$ we find by adding the equations that the total must be Gaussian
\begin{align}
    \tfrac{1}{\tau} \mathcal{L}_{a} (p_{\rm{a},\alpha} + p_{\rm{a},\beta}) &= 0.
\end{align}
We therefore obtain
\begin{align}
    p_{\rm{a},\alpha} + p_{\rm{a},\beta} &= \tfrac{1}{Z_a} e^{ - v(a) },
\end{align} 
such that we have exactly, and in all cases, $ p_{\rm{a},\beta} = \tfrac{1}{Z_a} e^{ - v(a) } - p_{\rm{a},\alpha} $. 
We are now able to obtain an equation in terms of the unknown density $p_{\rm{a},\alpha}(a)$ only, which takes the form
\begin{align}
    \tfrac{1}{\tau} \mathcal{L}_{a} p_{\rm{a},\alpha} - \big( R_{\alpha \beta} + R_{\beta \alpha} \big) p_{\rm{a},\alpha} + R_{\beta \alpha} \tfrac{1}{Z_a} e^{ - v(a) } = 0,
\end{align} 
As mentioned in Eq. \eqref{eq:ODEVeryLargePersistence}.
For symmetric wells and activity we have $R_{\beta \alpha}(a) = R_{\alpha \beta}(-a)$, from which it becomes useful to define $R_{\alpha \beta}(a) := R(a)$, resulting in $R_{\beta \alpha}(a) = R(-a)$. 

\subsubsection{First passage time distribution}
In the very high persistence time limit, where we have for the activity dependent rate $R(a) \approx R_0(a)$, we heuristically construct a time dependent equation, given some initial density $p_{\rm{a},\alpha}(a,t=t_0)$, by adding $R(a)$ as a sink term. We obtain
\begin{align}
    \partial_{t} p_{\rm{a},\alpha} = \frac{1}{\tau} \mathcal{L}_{a} p_{ \rm{a},\alpha } - R_{\alpha \beta}(a) p_{\rm{a},\alpha}.
    \label{eq:timedepedentverylargetau}
\end{align}
To calculate the First Passage Time Distribution (FPTD) in the very large $\tau$ limit, we need to take our initial density to be equal to the incoming probability from the other well in the steady state, such that $p_{\rm{a},\alpha}(a,t=t_0) = p_{\rm{a},\beta}(a) R_{\beta \alpha}(a)$. Then, integrating the sink term $- R_{\alpha \beta}(a) p_{\rm{a},\alpha}(a,t)$ over all activity $a$ will give the amount of probability leaving the well at any time $t$, thus giving the FPTD. We therefore have
\begin{align}
    {\rm{FPTD}} (t) = \int^{\infty}_{-\infty} R_{\alpha \beta}(a) p_{\rm{a},\alpha}(a,t) da,
\end{align}
given the initial condition $p_{\rm{a},\alpha}(a,t=0) = p_{\rm{a},\beta}(a) R_{\beta \alpha}(a)$.

\subsubsection{Infinite persistence time}
In the strict $\tau \rightarrow \infty$ limit, we have
\begin{align} 
    - R_{\alpha \beta}(a) p_{\rm{a},\alpha} + R_{\beta \alpha}(a) p_{\rm{a},\beta} = 0.
\end{align}
From this, we directly obtain the relationship
\begin{align}
    p_{\rm{a},\beta} = \frac{R_{\alpha \beta} }{R_{\beta \alpha} } p_{\rm{a},\alpha}.
    \label{eq:eff_detailed_balance}
\end{align} 
This is an effective detailed balance relation. This results form the fact that in the $\tau \rightarrow \infty$ limit the acitivity is completely static, such that the effective potential description, where $V(x) + \lambda a x$ is the effective potential, becomes exact for each value of $\alpha$.
We insert Eq. \eqref{eq:eff_detailed_balance} into the relation $p_{\rm{a},\alpha} + p_{\rm{a},\beta} = \tfrac{1}{Z_{a}} e^{-v(a)} $, from which we obtain
\begin{align}
    p_{\rm{a},\alpha}(a) = \frac{ \tfrac{1}{Z_{a}} e^{-v(a)} }{1 + \frac{ R_{\alpha \beta}(a) }{ R_{\beta \alpha}(a) } } 
\end{align}



\section{Intermediate regime} \label{Intermediate}

While we have working theories in the limits $\tau \ll 1/k$ and $\tau \gg 1/k$, there is no functioning theory around $\tau \sim 1/k$. Fortunately, this is a small inner region where the behaviour is constrained by the two asymptotic expansions that one has. Therefore, we will bridge the intermediate regime through a two point Padé approximation. We will bridge the the small persistence time regime to the intermediate-to-large persistence time regime, where one can assume that $p_{ \rm{a} } = \tfrac{1}{Z} e^{-v(a)}$, which leads to a constant value for $r(\tau)$. Although not necessary, this does make the two point Padé approximation simpler.

\subsection{Full Exponentiation of the Kramer's rate}
We begin with the Kramers--like expression for the escape rate,
\begin{equation}
r(\tau,\lambda)
\;\approx\;
\frac{D(x_{\min})}{2\pi}
\sqrt{\Phi''(x_{\min})\,\bigl|\Phi''(x_{\max})\bigr|}
\;\exp\!\bigl[\Phi(x_{\min}) - \Phi(x_{\max})\bigr],
\end{equation}
which we rewrite as
\begin{equation}
r(\tau,\lambda) = \exp\bigl[ A(\tau,\lambda) \bigr].
\end{equation}
Writing the sub-exponential factor in exponential form, and splitting the multiplication in the logarithm gives us
\begin{equation}
\label{eq:A-def}
A(\tau,\lambda)
=
\ln\!\left(
\frac{D(x_{\min})}{2\pi} \right) + \ln\!\left( \sqrt{\Phi''(x_{\min})\,\bigl|\Phi''(x_{\max})\bigr|}
\right)
+ \Phi(x_{\min}) - \Phi(x_{\max}).
\end{equation}

Throughout, $x_{\min}$ and $x_{\max}$ denote the positions of the minimum $x_{\text{min}} = -1$ and maximum $x_{\text{max}} = -1$ of the potential.  

\subsection{ Small persistence time expansion }

We expand all quantities appearing in \eqref{eq:A-def} in powers of $\tau$ at fixed~$\lambda$:
\begin{align}
\Delta \Phi &= \Delta \Phi_0 + \tau\,\Delta\Phi_1 + \tau^2\,\Delta\Phi_2 + \mathcal{O}(\tau^3),
\\[4pt]
D(x_{\min}) &= D_0 + \tau D_1 + \tau^2 D_2 + \mathcal{O}(\tau^3),
\\[4pt]
\Phi''(x_{\min}) &= \omega_0^- + \tau \omega_1^- + \tau^2 \omega_2^- + \mathcal{O}(\tau^3),
\\[4pt]
\Phi''(x_{\max}) &= \omega_0^+ + \tau \omega_1^+ + \tau^2 \omega_2^+ + \mathcal{O}(\tau^3),
\end{align}
where $\Delta\Phi_n := \Phi_n^+ - \Phi_n^-$ denotes the difference of the expansion coefficients of $\Phi$ at the minimum and maximum.
We expand the logarithmic parts as
\begin{align}
\ln\!\bigl(D_0 + \tau D_1 + \tau^2 D_2\bigr)
&= \ln D_0
+ \tau \frac{D_1}{D_0}
+ \tau^2\left(
\frac{D_2}{D_0}
- \frac{D_1^2}{2D_0^2}
\right)
+ \mathcal{O}(\tau^3),
\\[6pt]
\ln\!\bigl(\omega_0^\pm + \tau \omega_1^\pm + \tau^2 \omega_2^\pm \bigr)
&= \ln \omega_0^\pm
+ \tau \frac{\omega_1^\pm}{\omega_0^\pm}
+ \tau^2\left(
\frac{\omega_2^\pm}{\omega_0^\pm}
- \frac{(\omega_1^\pm)^2}{2 (\omega_0^\pm)^2}
\right)
+ \mathcal{O}(\tau^3).
\end{align}
The diffusion coefficients take the form
\begin{align}
    D_{0} &= 1, \\
    D_{1} &= \lambda^{2}, \\
    D_{2} &= - \lambda^{2} V''(x).
\end{align}
We evaluate the integral expression for $\Phi$ as
\begin{align}
    \int^{x} \frac{ F(u) }{ D(u) } du := - \Phi =  - \frac{ V(x) }{1+ \lambda^{2} \tau } - \lambda^{2} 
    \tau^{2} \frac{ V'(x)^{2}}{2(1+ \lambda^{2} \tau )^2},
\end{align}
and Taylor expand to find a polynomial in powers of $\tau$
\begin{align}
    \Phi = V + \tau \left( - \lambda^{2} V \right) + \tau^{2} 
    \left( \lambda^{4}  V + \frac{\lambda^{2}}{ 2 } V'^{2} \right) + \mathcal{O} \left( \tau^{3} \right). 
\end{align}
Substituting these expansions into \eqref{eq:A-def} and collecting powers of~$\tau$, we obtain
\begin{equation}
\label{eq:A-tau-expansion}
A(\tau,\lambda)
= A_0(\lambda) + \tau A_1(\lambda) + \tau^2 A_2(\lambda)
+ \mathcal{O}(\tau^3),
\end{equation}
where
\begin{align}
A_0(\lambda)
&= \ln\frac{D_0}{2\pi}
+ \frac12 \ln \omega_0^-
+ \frac12 \ln \bigl|\omega_0^+\bigr|
+ \Delta\Phi_0,
\\[4pt]
A_1(\lambda)
&= \frac{D_1}{D_0}
+ \frac12\!\left(
\frac{\omega_1^-}{\omega_0^-}
+ \frac{\omega_1^+}{\omega_0^+}
\right)
+ \Delta\Phi_1,
\\[4pt]
\label{eq:A2-general}
A_2(\lambda)
&= \frac{D_2}{D_0}
   - \frac{D_1^2}{2 D_0^2}
+ \frac12\!\left(
   \frac{\omega_2^-}{\omega_0^-}
   - \frac{(\omega_1^-)^2}{2 (\omega_0^-)^2}
   + \frac{\omega_2^+}{\omega_0^+}
   - \frac{(\omega_1^+)^2}{2 (\omega_0^+)^2}
  \right)
+ \Delta\Phi_2.
\end{align}
Accordingly, the Kramers rate in the small persistence time limit has the small--$\tau$ expansion
\begin{equation}
    r(\tau,\lambda)
    \;\approx\;
    \exp\!\Bigl[A_0(\lambda)
    + \tau A_1(\lambda)
    + \tau^2 A_2(\lambda)\Bigr],
\end{equation}

\subsection{Large persistence time expansion}

Because the rate varies over orders of magnitudes as $\tau$ changes, it instead becomes worthwhile to find the exponential dependence of the rate on $\tau$. Since in the large $\tau$ limit we asymptotically have $e^{ B_0 + B_1/\tau + B_2/\tau^2 + ...} = e^{B_0} [1+\frac{B_1}{\tau} + \frac{ B_2 + (1/2) B_1^{2} }{ \tau^{2} } + ... ]$, and since we only need the first correction $B_1$, it is sufficient to only acquire $b_1$ in Eq. \eqref{eq:rate_nonexp_largetau}. We find
\begin{align}
    r = \exp{ \left[ \ln{(b_0)} + \frac{b_1}{b_0} \frac{1}{\tau} + O \left( \frac{1}{\tau^2} \right) \right] }
\end{align}
It is this expression that we shall use for the two-point made approximation that bridges the intermediate regime.

\subsection{Pad\'e approximant} \label{Pade}

We have
\begin{align}
    r(\tau) = e^{R(\tau)},
\end{align}
where $R(\tau)$ is known only by its asymptotic
expansions in two opposite limits:
\begin{align}
R(\tau) &= A_0 + A_1 \tau + A_2 \tau^2 + O \left( \tau^{3} \right),
\qquad \tau \to 0, \label{eq:small-expansion}\\[4pt]
R(\tau) &= B_0 + \frac{B_1}{\tau} + O \left( \frac{1}{ \tau^{2} } \right),
\qquad \tau \to \infty, \label{eq:large-expansion}
\end{align}
where \(A_0,A_1,A_2,B_0,B_1\) are known.
We want to construct a rational function (a two--point Pad\'e
approximant) that reproduces these asymptotic behaviours up to the
orders shown.

We construct a Pad\'e approximant of type \([2/2]\), that is, a fraction of polynomials in $\tau$ that is equivalent in both asymptotes. We hypothesize that this will accurately bridge the intermediate behaviour. We then have
\begin{equation}
R(\tau) \approx
\frac{\alpha_0 + \alpha_1 \tau + \alpha_2 \tau^2}
     {1 + \beta_1 \tau + \beta_2 \tau^2},
\label{eq:pade-ansatz}
\end{equation}
where the denominator has been normalised so that its constant term is unity. We will determine the five unknown coefficients
\(\alpha_0,\alpha_1,\alpha_2,\beta_1,\beta_2\) by matching
\eqref{eq:pade-ansatz} to the Eqs.
\eqref{eq:small-expansion}--\eqref{eq:large-expansion} in the corresponding limits of $\tau$. The unique set of coefficients are obtained as
\begin{align}
\Delta &= A_0^2 - 2A_0B_0 + A_1 B_1 + B_0^2,
\\[4pt]
\alpha_0 &= A_0,\\[4pt]
\alpha_1 &=
\frac{-A_0 A_1 B_0 - A_0 A_2 B_1 + A_1^2 B_1 + A_1 B_0^2}{\Delta},
\\[4pt]
\alpha_2 &=
\frac{B_0(-A_0 A_2 + A_1^2 + A_2 B_0)}{\Delta},
\\[4pt]
\beta_1 &=
\frac{-A_0A_1 + A_1B_0 - A_2B_1}{\Delta},
\\[4pt]
\beta_2 &=
\frac{-A_0A_2 + A_1^2 + A_2 B_0}{\Delta}.
\end{align}
By construction, expanding $R(\tau)$ in powers of $\tau$ when $\tau \ll 1$ results in Eq. \eqref{eq:small-expansion}, while expanding $R(\tau)$ and $\frac{1}{\tau}$ results in the series in Eq. \eqref{eq:large-expansion}

\newpage

\section{Numerical methods}

For the numerical comparison of the transition rates we integrate the AOUP numerically. Path dynamics were integrated using a stochastic Runge-Kutta scheme of strong order 1.5 in the position $x$, while for the active variable $a$ the Euler–Maruyama method was employed. Using this integration scheme, we simulate two different cases for the $k=13$ barrier and the $k=7$ barrier. For the $k=13$ barrier, we have as the initial condition $x=-1$, while we draw activity $a$ out of the stationary Gaussian distribution $\frac{1}{\sqrt{2 \pi}} e^{-a^2 /2}$. We integrate until the particle crosses the boundary $x_c = 0.5$, thereby implementing an absorbing boundary, and repeat this process until we measure 20000 transitions. The rate is then determined by the mean transition time (MTT), such that $r = 1/\rm{MTT}$. For the $k=7$ barrier, we instead do not impose an absorbing boundary, but count when a particle has crossed from $x = -0.5$ to $x = 0.5$ or from $x = 0.5$ to $x = -0.5$. for the initial conditions we now have a probability of $1/2$ to start at $x=-1$ or $x=1$, while the initial $a$ is still drawn from the steady state distribution $\frac{1}{\sqrt{2 \pi}} e^{-a^2 /2}$. We let the system go to a steady state by waiting approximately 10 times the passive mean transition time, such that the counting only starts when $t > 10 \cdot {\rm MTT} = 10 \cdot 7 e^{-7}$.

We will now give the integration scheme. We denote the integration time step as $\Delta t$. At each step the random variables $\xi_1^n,\xi_2^n,\eta^n$ are generated, each drawn from a Gaussian distribution with variance $1$ and mean $0$. The stochastic increments are defined as
\begin{align}
\Delta W_n &= \sqrt{\Delta t}\,\xi_1^n, \\
I_{1,0}^{(n)} &= \sqrt{\frac{\Delta t^3}{3}}
\left(c\,\xi_1^n + \sqrt{1-c^2}\,\xi_2^n\right),
\end{align}
where $c = \sqrt{3}/2$ is a correlation coefficient that ensures that the covariance between $I^{(n)}_{1,0}$ and $\Delta W_n$ equals $\Delta t^2/2$. The position variable $x$ is updated as
\begin{align}
x_{n+1} = x_n
&+ (\lambda a_n + F(x_n))\Delta t
+ \sqrt{2}\,\Delta W_n
+ \sqrt{2}\,F'(x_n)\,I_{1,0}^{(n)} \\
&+ \frac{\Delta t^2}{2}
\left[
F(x_n)F'(x_n) +F''(x_n)
\right].
\end{align}
The activity value $a$ evolves according to the Ornstein-Uhlenbeck process and is integrated using the
Euler--Maruyama method,
\begin{equation}
a_{n+1} =
a_n - \frac{\Delta t}{\tau}a_n
+ \sqrt{\frac{2\Delta t}{\tau}}\;\eta^n .
\end{equation}


\end{document}